\documentstyle[12pt]{article}
\def\a{\alpha}  \def\d{\delta}
\def\b{\beta} \def\D{\Delta}

\def\g{\gamma}
\def\l{\lambda}
\def\o{\omega}
\def\vu{\eta}

\def\der{\frac{d}{du}}
\def\Res{\mbox{Res}_{u=x}}
\def\h{h\check{\:}}

\def\q2a{q^{2(\a ,\a)}}
\def\q-a{q^{-(\a ,\a)}}

\newcommand{\Rmt}{ universal $R$-matrix }
\newcommand{\bn}{\begin{equation}}
\newcommand{\ed}{\end{equation}}
\newtheorem{definition}{Definition}[section]
\newtheorem{proposition}{Proposition}[section]
\newtheorem{theorem}{Theorem}[section]
\newtheorem{lemma}{Lemma}[section]
\newtheorem{corollary}{Corollary}[section]

\newcommand{\hsp}{\mbox{$\hspace{.5in}$}}

\begin{document}
\title{{\bf
Yangian Double and Rational R-matrix}}
\author{ {\Large S.M. Khoroshkin}\thanks{e-mail
address:  khor@s43.msk.su}
\thanks{partially supported by AMS FSU foundation
and ISF grant MBI000}\\
Institute of Theoretical and Experimental Physics, \\
117259 Moscow, Russia\\ and \\{\Large
V.N.Tolstoy}\thanks{e-mail adress:
tolstoy@compnet.msu.su} \thanks{partially supported by
ISF grant MBI000} \\Institute of Nuclear
Physics, Moscow State University, \\119899 Moscow,
Russia } \date{} \maketitle
\begin{abstract}
Studying the algebraic structure of the double
${\cal D}Y(g)$ of the yangian $Y(g)$ we present the
triangular decomposition of ${\cal D}Y(g)$ and a
factorization for the canonical pairing of the yangian
with its dual inside ${\cal D}Y(g)$. As a
consequence we obtain an explicit formula for the
universal R-matrix $R$ of ${\cal D}Y(g)$ and demonstrate
how it works in evaluation representations of
$Y(sl_2)$.  We interprete one-dimensional factor
arising in concrete representations of $R$ as bilinear
form on highest weight polynomials of irreducible
representations of $Y(g)$ and express this form in
terms of {\it gamma-functions}.
\end{abstract}
\setcounter{footnote}{0}
\setcounter{equation}{0}
\section{Introduction}
Yangian $Y(g)$ of a simple Lie algebra $g$ was introduced by V.Drinfeld
\cite{D83} as a deformation of universal enveloping algebra $U(g[t])$ of
 a current algebra $g[t]$. The yangians $Y(g)$ and
quantum affine algebras $U_q(\hat{g})$ play the role
of dynamical symmetries in quantum field theories
\cite{BL} \cite{S}. Tensor products of
finite-dimensional representations of yangians produce
rational solutions of the Yang-Baxter equation;
 tensor products of
finite-dimensional representations of quantum affine
algebras  produce trigonometric solutions of the
Yang-Baxter equation. One can find out other deep
parallels in representation theories of yangians and
of quantum affine algebras. Nevertheless both of them
have their own original features.
 Yangian $Y(g)$ is much more closer to classical Lie
algebras, at least it contains universal enveloping
algebra $U(g)$ as a subalgebra; moreover the yangians
$Y(gl_n)$ could be defined entirely in terms of
classical representation theory \cite{O}.
The structure of quantum affine algebra $U_q(\hat{g})$
is more complicated. On the other hand, $U_q(\hat{g})$
inhabit the main properties of contragredient
algebras; Chevalley generators and $q$-deformed Serre
relations are permanent participants of the games with
quantum affine algebras.

The theory of Cartan-Weyl basis \cite{KT2}
\cite{KT3} allows to describe explicitely the
universal $R$-matrix, one of the main object in
physical applications. We have nothing of this for the
yangians, though suitable modifications of classical
methods of representation theory work for $Y(g)$ and
$U_q(\hat{g})$ as well (see \cite{Ch} \cite{D83}
\cite{J} \cite{NT} for instance). We want to make this
gap smaller.

It is more reasonable to work with quantum double
${\cal D}Y(g)$ of the yangian, if we take in mind physical
applications. The representations of ${\cal D}Y(g)$ do not
differ much from the representations of yangian
$Y(g)$: the extension of a representatin of $Y(g)$ to
representation of ${\cal D}Y(g)$ can be achieved just by
reexpansion of the currents from $Y(g)$ in other point
of projective line. We present here a study of
some algebraic properties of ${\cal D}Y(g)$, with the
accent to the canonical pairing in the double.

We prove that ${\cal D}Y(g)$ itself and a Hopf pairing of
$Y(g)$ with its dual inside ${\cal D}Y(g)$ admit triangular
decomposition analogous to Gauss decomposition of
ordinary matrices. This property gives possibility to
describe the pairing quite explicitely. As a
consequency we obtain explicite factorized expression
for the universal $R$-matrix of ${\cal D}Y(g)$ (completely
proved for ${\cal D}Y(sl_2)$ and partially in general case).
 To make the formulas more transparant, we present
 detailed calculations for ${\cal D}Y(sl_2)$, including the
 action of the universal $R$-matrix on evaluation
 representations.

  The most interesting factor $R_H$ of
 the universal $R$-matrix is concerned with (zero
 charge) Heisenberg subalgebra of ${\cal D}Y(g)$ which is a
 deformation of the currents to Cartan subalgebra $h$
 of $g$. Analogously to the case of $U_q(\hat{g})$
 \cite{KT2} \cite{KST} the structure of $R_H$ is
 governed by the $q$-analog of invariant scalar
 product in $h$; whenever $R_H$ acts on
 representations of ${\cal D}Y(g)$ a variable $q$ becomes a
shift operator $T:\; Tf(x)= f(x+1)$ (for quantum
affine algebras $q$ goes to multiplicative shift
$T_qf(x)= f(qx)$ in analogous situation).

After substitution of the universal $R$-matrix into
tensor product of concrete representations of ${\cal D}Y(g)$
we obtain more then usual rational $R$-matrix; an
additional information is concentrated in scalar phase
factor (scalar $S$-matrix) which we interprete as
bilinear multiplicative form on highest weight
polynomials of finite-dimensional representations
$V$ of $Y(g)$ (or, equivalently, on $K_0(Rep Y(g)$).
This form is a deformation of skewsymmetric form
$\frac{<,>}{a-b}$ on
irreducible evaluation representations of $g[t]$
 where $<,>$ is invariant scalar
product in $h^*$; $a$ and $b$ are the points where
evaluation representations are living. We present an
explicit expression of this form as some ratio of
$\Gamma$ functions defined by the structure of
$q$-analog of invariant scalar product in $h$.

 \bigskip
 {\bf Acknowlegements.} The authors are thankful to
 Professors V.Drinfeld, A.Molev, M.Nazarov,
 G.Olshanskii and A.Stolin for fruitful and
 stimulating discussions; they are also thankful to
 Mathematical Departement of Bielefeld University,
 where the first stages of the work were done, for
 kindful hospitality.

\setcounter{equation}{0}
\section{Yangian $Y(g)$ and its quantum double}
Let $g$ be a simple Lie algebra with Cartan matrix
$A_{i,j}$ and Chevalley generators $e_{i}=e_{\a_i}$,
$h_{i}=h_{\a_i}$, $f_{i}=f_{\a_i}=e_{-\a_i}$,
$i=1,\ldots,r$. Yangian $Y(g)$ is a deformation of the
 universal enveloping algebra $U(g[t])$. It can be
 defined \cite{D83}, \cite{Berkeley} as a Hopf
algebra with generators  $e_{i,k}=e_{\a_i,k}$,
$h_{i,k}=h_{\a_i,k}$, $f_{i,k}=f_{\a_i,k}$,
$i=1,\ldots ,r$, $\; k=0,1,\ldots$ subjected to
 relations
 $$[h_{i,k},h_{j,l}]=0,\hsp
  [h_{i,0}, e_{j,l}]= (\a_i,\a_j)e_{j,l},$$
  $$[h_{i,0}, f_{j,l}]= -(\a_i,\a_j)f_{j,l},\hsp
  [e_{i,k}, f_{j,l}]= \d_{i,j}h_{i,k+l}$$,
 \bn
 [h_{i,k+1}, e_{j,l}]-[h_{i,k},  e_{j,l+1}]=
  \frac{1}{2}(\a_i,\a_j)\{ h_{i,k},e_{j,l}\}
  \label{1.1}
  \ed
  where $\{ a,b\} =ab+ba$,
  $$[h_{i,k+1}, f_{j,l}]-[h_{i,k},  f_{j,l+1}]=
  -\frac{1}{2}(\a_i,\a_j)\{ h_{i,k},f_{j,l}\} $$
  $$[e_{i,k+1}, e_{j,l}]-[e_{i,k},  e_{j,l+1}]=
  \frac{1}{2}(\a_i,\a_j)\{ e_{i,k},e_{j,l}\}$$
  \bn
  [f_{i,k+1}, f_{j,l}]-[f_{i,k},  f_{j,l+1}]=
  -\frac{1}{2}(\a_i,\a_j)\{ f_{i,k},f_{j,l}\}
  \label{1.1a}
  \ed
\bn
 i\neq j,\;  n_{i,j}=1-A_{i,j}\longrightarrow\left\{
\begin{array}{l}
{\rm Sym}_{\{ k\} }[e_{i,k_1}[e_{i,k_2}\ldots
 [e_{i,k_{n_{i,j}}},e_{j,l}]\ldots ]]=0 \\\
{\rm Sym}_{\{ k\} }[f_{i,k_1}[f_{i,k_2}\ldots
 [f_{i,k_{n_{i,j}}},f_{j,l}]\ldots ]]=0
 \end{array}\right .
\label{1.1c}\ed

Subalgebra generated by $e_{i,0}$, $f_{i,0}$,
$h_{i,0}$, $i=1,2,\ldots ,r$ is naturally isomorphic
to $U(g)$ by (\ref{1.1})-(\ref{1.1a}). Using this
isomorphism we omit sometimes index zero for these
generators and consider the elements $e_\g$, $f_\g$,
$h_\g$, $\g\in\D_+(g)$ of Lie algebra $g$ as elements
of $Y(g)$. Here $\D_+(g)$ is a system of all positive
roots of $g$ and the root vectors $e_\g$, $f_\g$,
$h_\g$ are normalized by the condition $[e_\g ,f_\g
]=h_\g$.

One can deduce from original description of
$Y(g)$ \cite{D83} the following formulas of
comultiplication for basic generators of $Y(g)$:
\[
\D(x)=x\otimes 1+1\otimes x,\hsp x\in g,
\]
\[
\D(e_{i,1})=e_{i,1}\otimes 1+1\otimes
e_{i,1}+h_{i,0}\otimes e_{i,0}
-\sum_{\g\in\D_+(g)}f_\g\otimes
[e_{\a_i},e_\g],
\]
\[
\D(f_{i,1})=f_{i,1}\otimes 1+1\otimes
f_{i,1}+ f_{i,0}\otimes h_{i,0}+\sum_{\g\in\D_+(g)}
[f_{\a_i},f_\g]\otimes e_\g,
\]
\bn
\D(h_{i,1})=h_{i,1}\otimes 1+1\otimes
h_{i,1}-\sum_{\g\in\D_+(g)}(\a_i,\g)
f_\g\otimes e_\g
\label{1.2}
\ed

In this paper we are much more interested in quantum
double ${\cal D}Y(g)$ of $Y(g)$ (see \cite{Berkeley} for
definitions). The algebraic structure of ${\cal D}Y(g)$ can
be described as follows \cite{D83}.

Let $C(G)$ be an algebra generated by the elements
 $e_{i,k}$, $f_{i,k}$, $h_{i,k}$, $i=1,\ldots ,r$,
$k\in \mbox{{\bf Z}}$ subjected to relations
(\ref{1.1a})-(\ref{1.1c}). Algebra $C(g)$ admits {\bf
Z}-filtration
$$ \ldots\subset C_{n}\subset C_{n+1}\ldots \subset
C(g)$$
defined by the condition $\mbox{deg}\, e_{i,k}=
\mbox{deg}\, f_{i,k}=\mbox{deg}\, h_{i,k}=k$. Let
$\widehat{C}(g)$ be the corresponding formal
completion of $C(g)$. Actually Drinfeld \cite{D83}
proved that the double ${\cal D}Y(g)$ is isomorphic to
 $\widehat{C}(g)$ as an algebra. Generators
 $e_{i,k}$, $f_{i,k}$, $h_{i,k}$, $i=1,\ldots ,r$,
$k\geq 0$ define an inclusion $Y(g)\hookrightarrow
{\cal D}Y(g)$. We denote sometimes its image by $Y_+(g)$ or
 shortly by $Y$ when we need short notation. The dual
Hopf algebra with opposite comultiplication
$Y(g)^0$is isomorphic to $Y_-(g)=Y_-$ which is
generated by formal series $\sum_{k<0}a_k$,
$\mbox{deg}\, a_k=k$, and in this sense is generated
by the elements $e_{i,k}$, $f_{i,k}$, $h_{i,k}$,
$i=1,\ldots ,r$, $k<0$ (see also Remark to Theorem
 4.2).  To complete the description of ${\cal D}Y(g)$
 one should describe the structure of comultiplication
 in $Y_-(g)$ and the pairing between $Y_+(g)$ and
 $Y_-(g)$. This will be done below for $g=sl_2$ and
 partially for general case.

\setcounter{equation}{0}
\section{Triangular decomposition of ${\cal D}Y(g)$}
Let $E',H'$ and $F'$ be subalgebras without unit
element of $Y(g)$ generated by the elements $e_{i,k}$,
 $i=1,\ldots ,r$, $k\geq 0$; $h_{i,k}$, $i=1,\ldots
 ,r$, $k\geq 0$; $f_{i,k}$, $i=1,\ldots ,r$, $k\geq 0$
correspondingly. We denote also by $E,H$ and $F$ the
algebras $E',H'$ and $F'$ with added unit element.
 Following \cite{CP} one can deduce from (\ref{1.1a})-
(\ref{1.1c}) the following decomposition of $Y(g)$:
\begin{proposition}
\label{prop2.1}
 A multoplication in $Y(g)$ induces an isomorphism of
vector spaces
\bn
 E\otimes H\otimes F \leadsto\, Y(g)
\label{2.0}
\ed
\end{proposition}
We are going to
extend this decomposition to the double ${\cal D}Y(g)$
and factorize the natural pairing of $Y_+(g)$ and
$Y^0(g)\simeq Y_-(g)$ with respect to this
decomposition.  First we summarize the properties of
comultiplication in $Y=Y(g)$ which easily generalize
by induction the formulas (\ref{1.2}) (see also
\cite{CP}).  \begin{lemma}\label{prop2.2}\ \medskip

{\it (i)} For any $e\in E'$
\bn
\D(e)=e\otimes 1 \hsp \mbox{mod}\, Y\otimes E';
\label{2.1}
\ed

{\it (ii)} For any $f\in F'$
\bn
\D(f)=1\otimes f \hsp \mbox{mod}\, F'\otimes Y.
\label{2.2}
\ed
\end{lemma}
In particular, we conclude that $E$ is a right coideal
in $Y$ ($\D(E)\subset Y\otimes E$) and $F$ is a right
coideal ($\D(F)\subset F\otimes Y$) of $Y(g)$

Let also $\overline{E}$ be a subalgebra
(without unit element) of $Y(g)$ generated by the
elements $e_{i,k}$ and $h_{j,l}$, $i,j=1,\ldots ,r$,
$k,l\geq 0$;
 $\overline{F}$ be a subalgebra
(without unit element) of $Y(g)$ generated by the
elements $f_{i,k}$ and $h_{j,l}$, $i,j=1,\ldots ,r$,
$k,l\geq 0$.
We have also
\begin{lemma}\label{prop2.3}\ \medskip

{\it (i)} For any $e\in \overline{E}$
\bn
\D(e)=e\otimes 1 \hsp \mbox{mod}\, Y\otimes
\overline{E}; \label{2.3} \ed

{\it (ii)} For any $f\in \overline{F}$
\bn
\D(f)=1\otimes f \hsp \mbox{mod}\, \overline{F}\otimes
Y.  \label{2.4} \ed

{\it (iii)} For any $h\in H'$
\[
\D(h)=h\otimes 1 \;\;\;\;\mbox{mod}\,
Y\otimes\overline{E}\;=
\]
\bn
=1\otimes h
\;\;\;\;\mbox{mod}\, \overline{F}\otimes Y.
\label{2.5}
\ed
\end{lemma}
Let $<,>$ denotes the canonical Hopf pairing of
$Y_+(g)$=$Y(g)$ and its dual $Y^0(g)$. The Hopf
property of $<,>$ in this case can be read as $$
<a,cd>=<\D(a),c\otimes d>,\hsp <ab,c>=<b\otimes
a,\D(c)>$$
for any $a,b\in Y_+(g)$, and for any $c,d\in Y^0(g)$.
Here $<a\otimes b,c\otimes d>=<a,c><b,d>$.

Let now $E^*\subset Y^0(g)$ be defined as annulator
$(Y_+\overline{F})^\bot$ of $Y_+\overline{F}$, that is
\bn E^*=\{ e^*\in Y^0: \;
<y\overline{f},e^*>=0\;\;\forall y\in Y(g),\;\forall
\overline{f}\in \overline{F}\} .
\label{E}
\ed
Analogously, we define
\bn
F^*=(\overline{E}Y_+)^\bot, \hsp
\overline{E}^*=(Y_+F')^\bot,\hsp
\overline{F}^*=(E'Y_+)^\bot
\label{F}
\ed
and
\bn
H^*=(E'Y_+)^\bot\cap (Y_+F')^\bot .
\label{H}
\ed

The following lemma follows directly from the
definition of the Hopf pairing.
\begin{lemma}
Let $A$ and $A^*$ be two Hopf algebras with a Hopf
pairing $<,>: A\otimes A^*\rightarrow$ {\bf C}. Let a
subset $X\subset A$ satisfies the condition:
$$\D(X)\subset X\otimes A +A\otimes X.$$ Then both
 $(AX)^\bot$ and $(XA)^\bot$ are subalgebras of $A^*$.
\end{lemma}

\begin{corollary}
$E^*,\overline{E}^*, F,\overline{F}^*$ are subalgebras
of $Y^0(g)$.
\end{corollary}
The main result of this section is the following

\begin{theorem}
\label{th2.1}
For any $e\in E$, $h\in H$, $f\in F$; $e^*\in E^*$,
$h^*\in H^*$, $f^*\in F^*$ there is a factorization of
 canonical pairing:
\bn
<ehf,e^*h^*f^*>=<e,e^*><h,h^*><f,f^*>
\label{2.6}
\ed
\end{theorem}

As a consequence of Theorem \ref{th2.1} we have also
\begin{corollary}
A multiplication in $Y^0(g)$ induces an isomorphism of
vector spaces
\bn
E^*\otimes H^*\otimes F^* \leadsto\; Y^0(g)
\label{2.8}
\ed
\end{corollary}

The proof of Theorem \ref{th2.1} is strongly based on
the statements of Lemma \ref{prop2.2} and of
 Lemma \ref{prop2.3}. For instance, we check
first that
$$<e\overline{f},e^*\overline{f}^*>=<e,e^*>
<\overline{f},\overline{f}^*>$$
for any $e\in E'$, $e^*\in E^*$, $\overline{f}\in
\overline{F}$, $\overline{f}^*\in \overline{F}^*$.
Indeed, we have by definition
$$<e\overline{f}, e^*\overline{f}^*>=
=<\D(e)\D(\overline{f}), e^*\otimes\overline{f}^*>=$$
 \bn
= <(e\otimes 1+ \sum y_i \otimes e_i) (1 \otimes
\overline{f}+ \sum \overline{f}_j \otimes y_j), e^*
\otimes \overline{f}^*>
\label{2.9}
\ed
where $e_i\in E'$, $\overline{f}_j\in \overline{F}$ by
(\ref{2.1})-(\ref{2.4}). The rhs of (\ref{2.9}) is
equal to $<e \otimes \overline{f}, e^* \otimes
\overline{f}^*>$ by definition of $E^*$ and
$\overline{F}^*$.  Then we prove analogously that
$<eh, e^*h^*>=<e,e^*><h,h^*>$ for any $e\in E'$, $h\in
H'$, $e^*\in E^*$, $h^*\in H^*$ and then take  off
the primes.

Now we proceed with a more detailed study of the
pairing in yangian double. In the next section we
compute explicitely the pairing between  the
generators of $Y_+(g)$ and of $Y_-(g)\simeq Y^0(g)$.
\setcounter{equation}{0}
\section{Basic pairing for ${\cal D}Y(g)$}
The aim of this section is to compute explicitely the
pairing between generators $e_{i,k}$, $h_{i,k}$ and
$f_{i,k}$ of $Y_+(g)$ and of $Y_-(g)\simeq Y^0(g)$.
The answer will be written
in terms of generating functions (''fields'')
$e_{i,\pm} (u)$, $h_{i,\pm} (u)$ and $f_{i,\pm} (u)$
of $Y_\pm (g)$:  $$e_{i,+}(u)=\sum_{k\geq
0}e_{i,k}u^{-k-1},\hsp f_{i,+}(u)=\sum_{k\geq
0}f_{i,k}u^{-k-1},$$

$$h_{i,+}(u)=1+\sum_{k\geq
0}h_{i,k}u^{-k-1}$$ generate $Y_+(g)$ and
$$e_{i,-}(u)=-\sum_{k< 0}e_{i,k}u^{-k-1},\hsp
f_{i,-}(u)=-\sum_{k< 0}f_{i,k}u^{-k-1},$$

$$h_{i,-}(u)=1-\sum_{k< 0}h_{i,k}u^{-k-1}$$
generate $Y_-(g)$.

Explicit calculations will be done for the case of
 $Y(sl_2)$ where we omit for simplicity everywhere an
index $1$ of a simple root  (for instance, for generating
functions we use the notations $e_\pm (u)$, $h_\pm
(u)$ and $f_\pm (u)$).

The main result of this section for ${\cal D}Y(sl_2)$
may be formulated  in the following theorem:

\begin{theorem}
\label{th3.1}\
 {\it (i)} Subalgebras $E^*$, $H^*$ and $F^*$ (see
(\ref{E})-(\ref{H})) of $Y_-(sl_2)$ are generated
 by the fields $e_-(u)$, $h_-(u)$ and $f_-(u)$
 correspondingly;
 {\it (ii)} The pairing of the generators of
 $Y_\pm(sl_2)$ is given by the relations
 \bn
 <e_+(u),f_-(x)>=\frac{1}{u-x}\, ,\hsp
 <f_+(u),e_-(x)>=\frac{1}{u-x}\, ,
 \label{3.1}
 \ed
 \bn
 <h_+(u),h_-(x)>=\frac{u-x+1}{u-x-1}
 \hsp \mid x\mid \ll 1\ll\mid u\mid
 \label{3.2}
 \ed
 or, in terms of generators,
 $$<e_k,f_{-l}>=<f_k,e_{-l}>=-\d_{k,l-1},\hsp k\geq
 0,\; l>0,$$
 $$ <h_k,h_{-l}>= -2\frac{k!}{(l-1)!(k-l+1)!}\hsp
 k\geq 0,\;l>0$$
 \end{theorem}
 In general case we have analogous
\begin{theorem}
\label{th3.2}\
 {\it (i)} Subalgebras $E^*$, $H^*$ and $F^*$ (see
(\ref{E})-(\ref{H})) of $Y_-(g)$ are generated
 by the fields $e_{i,-}(u)$, $h_{i,-}(u)$ and
 $f_{i,-}(u)$ correspondingly ($i=1,\ldots ,r$);

{\it (ii)} The pairing of the generators of
 $Y_\pm(g)$ is given by the relations
 \bn
 <e_{i,+}(u),f_{j,-}(x)>=\frac{\d_{i,j}}{u-x}\, ,\hsp
 <f_{i,+}(u),e_{j,-}(x)>=\frac{\d_{i,j}}{u-x}\, ,\hsp
 \label{3.3}
 \ed
 \bn
 <h_{i,+}(u),h_{j,-}(x)>=\frac{u-x+\frac{1}{2}(\a_i,\a_j)}
 {u-x-\frac{1}{2}(\a_i,\a_j)}
\hsp \mid x\mid \ll 1\ll \mid u\mid
\label{3.4}
 \ed
 \end{theorem}

{\bf Remark.} As a corollary of Theorem \ref{th3.2} we
obtain a rigorous proof of Drinfeld's Theorem,
mentioned in Section 2: the double ${\cal D}Y(g)$ is
isomorphic to an algebra $\widehat{C}(g)$, in
  particular, dual to $Y(g)$ Hopf algebra with
  opposite comultiplication $Y^0(g)$ is
  isomorphic to $Y_-(g)$.
  \bigskip

 We can describe explicitely the yangian $Y(sl_2)$ and
  its quantum double in terms of generating
 functions. One can check that the defining relation
 (\ref{1.1})-(\ref{1.1c}) for $Y(sl_2)$ are
 equivalent to the following conditions on $e_+(u)$,
 $h_+(u)$, $f_+(u)$:
 $$ [h_+(u),h_+(v)]=0$$ $$
 [e_+(u),e_+(v)]=\frac{(e_+(v)-e_+(u))^2}{v-u},\hsp
[f_+(u),f_+(v)]=\frac{(f_+(u)-f_+(v))^2}{u-v},$$
 $$ [h_+(u),e_+(v)]=\frac{\{ h_+(u),e_+(v)-e_+(u)\}
 }{u-v},$$
 $$ [h_+(u),f_+(v)]=\frac{\{
 h_+(u),f_+(u)-f_+(v)\} }{u-v},$$
\bn
 [e_+(u),f_+(v)]=\frac{h_+(u)-h_+(v)}{u-v}.
\label{3.3a}
\ed
 Moreover, A.I.Molev \cite{Molev} showed that the
 comultiplication in $Y(sl_2)$ can be described as
 follows:
 $$ \D(e_+(u))= e_+(u)\otimes 1 +\sum_{k=0}^\infty
 (-1)^kf_+^k(u+1)h_+(u)\otimes e_+^{k+1}(u)=$$
 \bn
 =e_+(u)\otimes 1 +\sum_{k=0}^\infty
 (-1)^kh_+(u)f_+^k(u-1)\otimes e_+^{k+1}(u),
 \label{3.4a}
 \ed
 $$ \D(f_+(u))= 1\otimes f_+(u)
 +\sum_{k=0}^\infty (-1)^ke_+^{k+1}(u)\otimes
 h_+(u)e_+^{k}(u+1)=$$ \bn =1\otimes f_+(u)
 +\sum_{k=0}^\infty (-1)^ke_+^{k+1}(u)\otimes
 e_+^{k}(u-1)h_+(u)
  \label{3.4b}
 \ed
 $$ \D(h_+(u))= \sum_{k=0}^\infty
 (-1)^k(k+1)f_+^k(u+1)h_+(u)\otimes
h_+(u)e_+^{k+1}(u+1)=$$
 \bn =
 \sum_{k=0}^\infty (-1)^k
 (k+1)h_+(u)f_+^k(u-1)\otimes
 e_+^{k+1}(u-1)h_+(u),
 \label{3.4c}
 \ed

 The generators of ${\cal D}Y(sl_2)$ satisfy the relations
 $$ [h_\pm(u),h_\pm(v)]=0,\hsp
 [h_+(u),h_-(v)]=0$$
$$[e_\pm(u),e_\pm(v)]=\frac{(e_\pm(v)-e_\pm(u))^2}{v-u},
\hsp
[f_\pm(u),f_\pm(v)]=\frac{(f_+(u)-f_+(v))^2}{u-v},$$
$$ [h_\pm(u),e_\pm(v)]=\frac{\{
 h_\pm(u),e_\pm(v)-e_\pm(u)\} }{u-v},$$ $$
 [h_\pm(u),f_\pm(v)]=\frac{\{
 h_\pm(u),f_\pm(u)-f_\pm(v)\} }{u-v},$$
 \[
 [e_\pm(u),f_\pm(v)]=\frac{h_\pm(u)-h_\pm(v)}{u-v},
\]
$$ [e_+(u),e_-(v)]=\frac{(e_+(v)-e_-(u))^2}{v-u},\hsp
  [f_+(u),f_-(v)]=\frac{(f_+(u)-f_-(v))^2}{u-v},$$
 $$ [h_+(u),e_-(v)]=\frac{\{ h_+(u),e_-(v)-e_+(u)\}
 }{u-v},$$
 $$ [h_-(u),e_+(v)]=\frac{\{ h_-(u),e_+(v)-e_-(u)\}
 }{u-v},$$
 $$ [h_+(u),f_-(v)]=\frac{\{
 h_+(u),f_+(u)-f_-(v)\} }{u-v},$$
 $$ [h_-(u),f_+(v)]=\frac{\{
 h_+(u),f_-(u)-f_+(v)\} }{u-v},$$
\bn
 [e_+(u),f_-(v)]=\frac{h_-(v)-h_+(u)}{u-v}, \hsp
 [e_-(u),f_+(v)]=\frac{h_+(v)-h_-(u)}{u-v}.
\label{3.3ab}
\ed
For the comultiplication in $Y_-(g)$ one can use
formulas (\ref{3.4a}), (\ref{3.4b}) with subindex $+$
replaced everywhere by $-$.

The rest of the section is devoted to the proof of the
Theorem \ref{th3.1}. (the proof of Theorem \ref{th3.2}
is quite analogous).

For the proof of Theorem \ref{th3.1} we need a
stronger version of Lemmas \ref{prop2.2} and
\ref{prop2.3} for $Y=Y(sl_2)$. One can find them in
\cite{CP} or deduce directly from
(\ref{3.4a})-(\ref{3.4c}). In the notations of the
previous section we have
 \begin{proposition}\cite{CP}
\label{prop3.1} The following properties of
comultiplication are valid for $Y(sl_2)$:
$$\D(e_+(u))=e_+(u)\otimes 1+h_+(u)\otimes e_+(u)
\;\;\; {\rm mod}\; YF\otimes E,$$
\bn
\D(f_+(u))=1\otimes f_+(u)+f_+(u)\otimes h_+(u) \;\;\;
{\rm mod}\; F\otimes EY,
\label{3.6}
\ed

$$\D(h_+(u))=h_+(u)\otimes h_+(u)
\;\;\; {\rm mod}\; YF\otimes EY.$$
\end{proposition}
We can use rhs of (\ref{3.6}) in order to compute for
instance the terms of $\D( e_+(u)e_+(v))$ which give
nonzero contribution to the pairing with elements of
$E^*\otimes E^*$. More precisely, using commutation
relations (\ref{3.3}) we have:
$$\D( e_+(u)e_+(v))=e_+(u)e_+(v)\otimes 1+1\otimes
 e_+(u)e_+(v) +$$
\bn
+\frac{u-v+1}{u-v-1}e_+(v)\otimes
e_+(u)-\frac{2}{u-v-1}e_+(u-1)\otimes e_+(u)\hsp {\rm
mod}\; YF\otimes E,
\label{3.7}
\ed

$$\D( f_+(u)f_+(v))=f_+(u)f_+(v)\otimes 1+1\otimes
 f_+(u)f_+(v)) +$$

\bn
+\frac{u-v-1}{u-v+1}f_+(v)\otimes
 f_+(u)-\frac{2}{u-v+1}f_+(v)\otimes f_+(v-1)\hsp {\rm
mod}\; F\otimes EY,
\label{3.8}
\ed

$$\D( h_+(u)f_+(v))=
h_+(u)\otimes h_+(u)f_+(v)+
h_+(u)f_+(v)\otimes h_+(u)h_+(v),$$

\bn
\D( e_+(u)h_+(v))=
e_+(u)h_+(v)\otimes h_+(v)+
h_+(u)h_+(v)\otimes e_+(u)h_+(v)
\;\;\;\;{\rm mod}\; YF\otimes EY
\label{3.10}
\ed\medskip

Let now $e^*(x)$ be a generating function of some
elements $e_i^*\in E^*$,
$e^*(x)=\sum_{i<0}e_i^*x^{-i-1}$ such that the pairing
 $<e_+(u),e^*(x)>=E(x-u)$ , $\mid x\mid \ll 1\ll\mid
u\mid$ depends on $u-x$ only. Analogously, let
 $f^*(x)$ be a generating function of some
elements $f_i^*\in F^*$,
$f^*(x)=\sum_{i<0}f_i^*x^{-i-1}$ such that the pairing
 $<f_+(u),f^*(x)>=F(x-u)$ , $\mid x\mid \ll 1\ll\mid
 u\mid$ depends on $u-x$ only. Using (\ref{3.6}),
(\ref{3.7}) one can prove the following
\begin{proposition}
\label{prop3.2}\
 {\it (i)} The conditions
\bn
<e_+(u),e^*(x)>=E(u-x)=\frac{\a}{u-x}\hsp \mbox{for
some }\:\a\in \mbox{{\bf C}}
\label{3.11}
\ed
and
\bn
[e^*(x),e^*(y)]=\frac{(e^*(x)-e^*(y)^2}{x-y}
\label{3.12}
\ed
are equivalent;

{\it (ii)} the conditions
\bn
<f_+(u),f^*(x)>=F(u-x)=\frac{\b}{u-x}\hsp \mbox{for
some }\:\b\in \mbox{{\bf C}}
\label{3.13}
\ed
and
\bn
[f^*(x),f^*(y)]=\frac{(f^*(y)-f^*(x)^2}{x-y}
\label{3.14}
\ed
are equivalent;
\end{proposition}
Let now $h^*(x)=1+\sum_{i<0}h_i^*x^{-i-1}$ be a
generating function of some elements $h_i^*\in H^*$
such that \\
{\it (i)} a pairing $<h_+(u),h^*(x)>=H(u-x)$ depends
on $u-x$ only;\\ {\it (ii)} the pairing $H(u-x)$ is
multiplicative on $h_+(u)$:
$<h_+(u)h_+(v),h^*(x)>=H(v-x)H(u-x)$ (the last
condition is natural due to multiplicative structure
of $\D(h_+(u))$ mod $YF\otimes EY$).

The calculations analogous to ones from Proposition
\ref{prop3.2} demonstrate the following
kharacterization of the pairing $H(u-x)$:
\begin{proposition}
\label{prop3.3}\
{\it (i)} Let $e^*(x)$ satisfies the conditions of
Proposition \ref{prop3.2}. Then the equalities
\bn
<h_+(u),h^*(x)>=\frac{u-x+1}{u-x-1}
\label{3.15a}
\ed
and
\bn
[h^*(x),e^*(y)]=\frac{\{ h^*(x),e^*(y)-e^*(x)\} }{x-y}
\label{3.16a};
\ed

{\it (ii)} Let $f^*(x)$ satisfies the conditions of
Proposition \ref{prop3.2}. Then the equalities
\bn
<h_+(u),h^*(x)>=\frac{u-x+1}{u-x-1}
\label{3.15}
\ed
and
\bn
[h^*(x),f^*(y)]=\frac{\{ h^*(x),f^*(x)-f^*(y)\} }{x-y}
\label{3.16};
\ed
are equivalent.
\end{proposition}
Analogously, one can see that under the conditions of
Proposition \ref{prop3.2} and of Proposition
\ref{prop3.3} the following relation is valid:
\bn
[f^*(x),e^*(y)]=\a\b\frac{ h^*(x)-h^*(y)}{x-y}.
 \label{3.19}
\ed
Comparing (\ref{3.11})-(\ref{3.19}) with (\ref{3.3ab})
we conclude that $\a\b =1$ and we can identify
$h^*(x)$ with $ h_-(x)$,
 $e^*(x)$ with $\g f_-(x)$ and $f^*(x)$ with $\g^{-1}
e_-(x)$
for some $\g\in \mbox{{\bf C}}^*$. We should find the
constant $\g$ to make the proof of Theorem \ref{th3.1}
 complete. Such an information can be extracted from
Yang $R$-matrix $R=1+\frac{P}{a-b}$ acting in tensor
product $V(a)\otimes V(b)$ of two-dimensional
representations of ${\cal D}Y(sl_2)$. The action of
generators of ${\cal D}Y(sl_2)$  in $V(c)$ wih a basis
$v_1$, $v_2$ can be described by the formulas:
$$e_i(v_1)=0,\;\;e_i(v_2)=c^iv_1,\;\;
f_i(v_2)=0,\;\;f_i(v_1)=c^iv_2,\;\;,$$
 $$h_i(v_1)=c^iv_1,\;\;h_i(v_2)=-c^iv_2,\;\;,$$
According to Theorem \ref{th2.1} (see
 also further Proposition \ref{prop4.1}, the
 reformulation in terms of the universal $R$-matrix)
 we take Gauss decomposition of Yang $R$-matrix:
 $R=R_ER_HR_F$ and find that
 $$R_E=1+\frac{1}{a-b}e_0\otimes f_0=1-\sum_{i\geq
 0}e_i\otimes f_{-i-1}$$
 which gives $<e_i,f_{-i-1}>=-1$, $i\geq 0$ or,
 equivalently $<e_+(u),f_-(x)>= \frac{1}{u-x}$. Thus
$\g =1$ which complete the proof of Theorem
 \ref{th3.1}.

 {\bf Remarks. 1.} Actually a variant of pairing
 (\ref{3.3})-(\ref{3.4}) was computed by Drinfeld
 \cite{Dunpublished}. It appeared as one of the basic
 points of his quantization of $g[t]$.

 {\bf 2.} The pairing  (\ref{3.3})-(\ref{3.4}) may be
 considered as a deformation of the classical pairing
 in $g[[t^{-1},t]$ given by rational $r$-matrix
 $r=\frac{\Omega}{u}$ where $\Omega$ is divided
 Casimir operator. Formulas (\ref{3.3})-(\ref{3.4})
 show that this pairing remains unchanged for the
 currents to nilpotent subalgebras and changes by
 shifts $\pm \frac{1}{2}(\a_i,\a_j)$ in (de)nominators
 of the pairing functions  of the current to Cartan
 subalgebras.
\setcounter{equation}{0}
\section{The universal $R$-matrix for ${\cal D}Y(g)$}
Let us remind that the universal $R$-matrix
\cite{Berkeley} for a quasitriangular Hopf algebra
${\cal A}$ is an invertible element $R$ of some
extension of ${\cal A}\otimes {\cal A}$ satisfying the
conditions
\bn
{\D}'(x)=R\D(x)R^{-1}\hsp \forall x\in {\cal A},
\label{4.0a}
\ed
\bn
(\D\otimes id)R=R^{13}R^{23},\hsp (id\otimes
\D)R=R^{13}R^{12}
\label{4.0}
\ed
where ${\D}'=\sigma \D$, $\sigma(u\otimes v)=v\otimes
u$ is an opposite comultiplication in ${\cal A}$. If
${\cal A}$ is a quantum double of a Hopf algebra
${\cal A}_+$, ${\cal A}  \simeq {\cal A}_+\otimes
{\cal A}_-$, ${\cal A}_-  $ being dual to ${\cal A}_+$
with an opposite comultiplication, then ${\cal A}$
admits a canonical presentation of the universal
$R$-matrix $R=\sum e_i\otimes e^i$, where $e_i$ and
$e^i$ are dual bases of ${\cal A}_+$ and of
 ${\cal A}_-$.

In the case of yangian ${\cal A}_+$ is $Y(g)$ and $R$
is a canonical universal $R$-matrix in ${\cal D}Y(g)$. In the
notations of section 3 we have, due to
part (i) of Theorem \ref{th3.2} the following
reformulation of Theorem \ref{th2.1}.
\begin{proposition}
\label{prop4.1}
Let $E_\pm$, $H_\pm$, $F_\pm$ be subalgebras of
${\cal D}Y(g)$ generated by $e_{i,\pm}$, $h_{i,\pm}$,
$f_{i,\pm}$, $i=1,\ldots ,r$. Then the universal
$R$-matrix $R$ of ${\cal D}Y(g)$ can be factorized as
\bn
R=R_ER_HR_F
\label{4.1}
\ed
where $R_E\in E_+\otimes E_-$,
$R_H\in H_+\otimes H_-$, $R_F\in F_+\otimes F_-$.
\end{proposition}
Explicit expressions of $R_E$ and $R_F$  for
${\cal D}Y(sl_2)$ can be found quite easily by inductive
computation of the pairing for $E_+$ and $E_-$,
$F_+$ and $F_-$ (compare \cite{Ro}):
\[
<e_0^{n_0}e_1^{n_1}\cdots e_k^{n_k},
f_{-1}^{m_0}f_{-2}^{m_1}\cdots f_{-k-1}^{m_k}>=
\]
\bn
=(-1)^{n_0+\ldots n_k}
\d_{n_0,m_0}\d_{n_1,m_1}\cdots \d_{n_k,m_k}\cdot
n_0!n_1!\cdots n_k!,
\label{4.2}
\ed
\[
<f_k^{n_k}\cdots f_1^{n_1} f_0^{n_0},
e_{-k-1}^{m_k}\cdots e_{-2}^{m_1}\cdots
e_{-1}^{m_0}>= \] \bn =(-1)^{n_0+\ldots n_k}
\d_{n_0,m_0}\d_{n_1,m_1}\cdots \d_{n_k,m_k}\cdot
n_0!n_1!\cdots n_k!,
\label{4.3}
\ed \medskip
 As an immediate corollary of (\ref{4.2})-(\ref{4.3})
we have
\begin{theorem}
\label{th4.1}
The factors $R_E$ and $R_F$ of the universal
$R$-matrix for ${\cal D}Y(sl_2)$ can be written as
\bn
R_E=\; \prod_{i\geq 0}^{\rightarrow}
\exp (-e_i\otimes f_{-i-1})=\;\;
\exp (-e_0\otimes f_{-1})  \cdot
\exp (-e_1\otimes f_{-2})  \cdot \ldots ,
\label{4.4}
\ed
\bn
R_F=\; \prod_{i\geq 0}^{\leftarrow}
\exp (-f_i\otimes e_{-i-1})=\;\;
\ldots \cdot \exp (-f_1\otimes e_{-2})  \cdot
\exp (-f_0\otimes e_{-1}).
\label{4.5}
\ed
\end {theorem}
For the general case of Theorem \ref{th4.1} see the
end of this section (formulas (\ref{4.30}),
(\ref{4.31})).
%%%%%%%%%%%%%%%%%%%%%input reference%%%%%%%%%%%%%%

The middle term $R_H$ of the universal $R$-matrix $R$
of ${\cal D}Y(g)$ has more delicate structure. One can find
it directly after huge calculations but we prefer to
use more elegant arguments of the connection between
two realizations of ${\cal D}Y(g)$. The general scheme is as
follows. Let $\widetilde{{\cal D}Y}(g)$ be a Hopf algebra
isomorphic to ${\cal D}Y(g)$ as an algebra with a following
comultiplication \cite{Dnew}, which naturally appears
in a quantization of current algebra $g[t]$:
\[
\tilde{\D}(e_i(u))=e_i(u)\otimes 1+h_{i,-}(u)\otimes
e_i(u),
\]
\[
\tilde{\D}(f_i(u))=1\otimes f_i(u)
+f_i(u)\otimes h_{i,+},
 \]
\bn
\tilde{\D}(h_{i,\pm}(u))=
h_{i,\pm}(u)\otimes h_{i,\pm}(u)
\label{4.6}
 \ed
where $e_i(u)=e_{i,+}(u)-e_{i,-}(u)$,
 $f_i(u)=f_{i,+}(u)-f_{i,-}(u)$ or
$e_{i}(u)=\sum_{n\in Z}e_{i,n}u^{-n-1}$,
$f_{i}(u)=\sum_{n\in Z}f_{i,n}u^{-n-1}$.

The arguments of \cite{bonn} show that just as for
$U_q(\hat{g})$ the Hopf algebra structure of
$\widetilde{{\cal D}Y}(g)$  is connected with a Hopf algebra
structure of ${\cal D}Y(g)$ via twisting of comultiplication
by action of the longest (virtual) element $w_0$ of
affine Weyl group. The elements of $H_{\pm}$ are
stable under this action thus the pairing
$<h_{i,+}(u),h_{j,-}(x)>$ is the same in ${\cal D}Y(g)$ and in
$\widetilde{{\cal D}Y}(g)$.  Formulas (\ref{4.6}) show that
the components of $K_{i,\pm}(u)= \log h_{i,\pm}(u)$
are primitive elements of $\widetilde{{\cal D}Y}(g)$.
This allows us to write down immediately the pairing
for the whole subalgebras $H_\pm$; the computation of
the factor $R_H$ reduces to diagonalization of the
form $<K_{i,+}(u), K_{j,-}(x)>$. Explicit
diagonalization will be done later, now we formulate a
general statement about the structure of $R_H$.
\begin{theorem}
\label{th4.2}
Let $K_{i,\pm}(u)= \log h_{i,\pm}(u)=
\sum k_{i,n,\pm}u^{-n-1}$. Let $\Psi$ be a linear
space generated by  the elements
$k_{i,n,+}$; $\Phi$ be a linear space generated by
the elements $k_{i,n,-}$. Then the factor $R_H$ of the
universal $R$-matrix for ${\cal D}Y(g)$ has a form
\bn
R_H=\exp \sum_a \psi_a\otimes\phi^a
\label{4.14}
\ed
where $\sum_a \psi_a\otimes\phi^a$ is a canonical
tensor in $\Psi\otimes\Phi$ with respect to the
pairing
\bn
<K_{i,+}(u),K_{j,-}(x)>=
\log\frac{u-x+\frac{1}{2}(\a_i,\a_j)}
{u-x-\frac{1}{2}(\a_i,\a_j)}, \hsp
\mid x\mid \ll 1\ll\mid u\mid
 \label{4.15}
\ed
\end{theorem}
\medskip {\bf Proof}
%%of Theorem \ref{th3.2}.\\
There is no action of affine Weyl group on ${\cal D}Y(g)$:
natural analogs of simple reflections map ${\cal D}Y(g)$ into
a different algebra. Nevertheless the affine shifts in
${\cal D}Y(g)$ are well defined. Let, for instance,
$\overline{w}_0$ be the following automorphism of
${\cal D}Y(g)$:
\bn
\overline{w}_0(e_{i,n})=e_{i,n+1},\hsp
\overline{w}_0(f_{i,n})=f_{i,n-1},\hsp
\overline{w}_0(h_{i,n})=h_{i,n}.
\label{4.5a}
\ed
The arguments of \cite{bonn} applied to ${\cal D}Y(g)$ give
the following
\begin{proposition}

\label{prop4.3} A Hopf algebra $\widetilde{{\cal D}Y}(g)$ is
isomorphic to ${\cal D}Y(g)$ with a comultiplication twisted
by $w_0=\lim_{n\rightarrow\infty}\overline{w}_0^n$.
 In other words,  for any $x\in {\cal D}Y(g)$
\bn
\tilde{\D}(x)=\D^{w_0}(x):=
\lim_{n\rightarrow\infty}\overline{w}_0^{\,n}\otimes
\overline{w}_0^{\,n}\D(\overline{w}_0^{\,-n}x)
\label{4.6a}
\ed
for a suitable topology in ${\cal D}Y(g)\otimes {\cal D}Y(g)$ (see
\cite{bonn}).
 \end{proposition}
 The Hopf algebra $\widetilde{{\cal D}Y}(g)$ is by definition
 a double of $\widetilde{{\cal D}Y}_+(g)$ where
 $\widetilde{{\cal D}Y}_+(g)$ is generated by
 $e_{i,n}$,
 $n\in$ {\bf Z} and $h_{i,n}$, $n\geq 0$. Let
 $\widetilde{{\cal D}Y}_-(g)$ be generated by $f_{i,n}$,
 $n\in$ {\bf Z} and $h_{i,n}$, $n< 0$. Then
 $\widetilde{{\cal D}Y}_-(g)$ is isomorphic to a dual
of $\widetilde{{\cal D}Y}_+(g)$ with an opposite
 comultiplication. Proposition \ref{prop4.3} allows
 one to compute the pairing $\widetilde{{\cal
 D}Y}_+(g)\otimes \widetilde{{\cal
 D}Y}_+(g)\rightarrow$ {\bf C}. Before computing this
 pairing let us say first some general words about
 Hopf pairings and automorphisms.

 Let $A$ and $B$ be two Hopf algebras with a Hopf
 pairing $<,>: A\otimes B \rightarrow$ {\bf C}.  Let
 $w_A: \: A\rightarrow A'$ and $w_B: \: B\rightarrow
 B'$ be some isomorphisms of algebras. Then the
 algebras $A'$ and $B'$ can be canonically equipped
 with a structure of Hopf algebras if we define
 comultiplications $\D^{w_A}:A'\rightarrow A'$ and
 $\D^{w_B}:B'\rightarrow B'$ by the rules
 \bn
 \D^{w_A}(a')=w_A\otimes w_A \D(w_A^{-1}a'),\hsp
 \D^{w_B}(b')=w_B\otimes w_B \D(w_B^{-1}b').
 \label{4.7}
 \ed
 Moreover, the pairing
 \bn
 <,>:A'\otimes B'\rightarrow \mbox{{\bf C}},\hsp
 <a',b'>=<w_A^{-1}a',w_B^{-1}b'>
 \label{4.8}
 \ed
 is a Hopf pairing.

 Taking $A=Y_+(g)$, $B=Y_-(g)$,
 $w_A= w_B= \overline{w}_0^n$ we obtain a Hopf pairing
 between $Y_+^n(g)$ and  $Y_-^n(g)$ where $Y_+^n(g)=
 \overline{w}_0^nY_+(g)$, $Y_-^n(g)=
 \overline{w}_0^nY_-(g)$. One can easily see from
 (\ref{3.3}), (\ref{3.4}) that this pairing stabilizes
 when $n\rightarrow \infty$ and defines a Hopf pairing
 between  $\widetilde{{\cal D}Y}_+(g)$ and
 $\widetilde{{\cal D}Y}_-(g)$ (which should actually coincide
 with that from a double structure of
 $\widetilde{{\cal D}Y}(g)$). This pairing looks like
 \bn
 <e_{i,n}, f_{j,-m-1}>=-\d_{i,j}\d_{m,n},
 \label{4.9}
 \ed
 \bn
 <h_+(u),h_-(x)>=\frac{u-x+\frac{1}{2}(\a_i,\a_j)}
  {u-x-\frac{1}{2}(\a_i,\a_j)}, \hsp
\mid x\mid \ll 1\ll \mid u\mid
 \label{4.10}
 \ed
The pairing (\ref{4.10}) coincides with (\ref{3.4}) because the
elements $h_{i,n}$ remain stable under the action of automorphism
 $\overline{w}_0$. Let $K_{i,\pm}(u)= \log h_{i,\pm}(u)$. Then the
  relation (\ref{4.3}) show that
  \bn
  \tilde{\D}(K_{i,\pm}(u))= K_{i,\pm}(u)\otimes 1 +1\otimes
  K_{i,\pm}(u),
  \label{4.11}
  \ed
  \bn
  <K_{i,+}(u),K_{i,-}(x)>= \log
  \frac{u-x+\frac{1}{2}(\a_i,\a_j)}
  {u-x-\frac{1}{2}(\a_i,\a_j)}, \hsp
\mid x\mid \ll 1\ll \mid u\mid
  \label{4.12}
  \ed
  which means that the coefficients of $K_{i,\pm}(u)$ are primitive
  elements with respect to $\tilde{\D}$ (an element $a$ is primitive
  with respect to $\tilde{\D}$ if $\tilde{D}(a)= a\otimes 1 + 1 \otimes
  a$).
  Now we are in the condition of the following simple general
  statement.

  Let $A$ and $B$ be two dual Hopf algebras isomorphic as algebras to
  free commutative algebras $A\simeq \mbox{{\bf
C}}[\Psi ]$, $B\simeq\mbox{{\bf C}}[\Phi ]$ where
  $\Psi $ and $\Phi$ are vectorspaces of generators,
  such that all $\psi\in \Psi$ (or all $\phi\in \Phi$)
  are primitive elements. Let $\psi_a$ and $\phi^a$
  be dual bases of $\Psi$ and $\Phi$ with respect to
  (nondegenerated) restriction of the pairing to
  $\Psi\otimes\Phi$ .  Then, once we choose some order
  of basic vectors, we have \[
  <\psi_{a_{1}}^{n_1}\cdots \psi_{a_{k}}^{n_k},
 (\phi^{a_{1}})^{m_1}\cdots (\phi^{a_{l}})^{m_l}>=
 \d_{k,l}\d_{n_1,m_1}\cdots \d_{n_k,m_k}
 n_1!\cdots n_k!
 \]
 or, in other words, the canonical tensor $R_{A\otimes B}= \sum a_i
  \otimes b^i$ of the pairing of $A$ and $B$ is
 an exponential of the canonical tensor $\Omega= \sum
  \psi_a\otimes\phi^a$ of the pairing of $\Psi$ and
  $\Phi$:
\bn
R_{A\otimes B}=\exp \Omega
\label{5.18a}
  \ed
which  proves Theorem \ref{th4.2}.

  We describe further the canonical tensor
  $\Omega= \sum \psi_a\otimes\phi^a$ from  Theorem \ref{th4.2} more
  explicitly, in other words we present concrete diagonalization of
  bilinear form (\ref{4.15}). For illustration, we first do this for
  ${\cal D}Y(sl_2)$.

  We have from (\ref{4.15})
  \bn
  <\der K_+(u),K_-(x)>=\frac{1}{u-x+1}-\frac{1}{u-x-1}
  \label{4.16}
  \ed
If $\psi(u)=\sum_{i\geq 0}\psi_iu^{-i-1}$,
 $\phi(x)=\sum_{i< 0}\phi_ix^{-i-1}$
 are arbitrary fields from $\Psi$ and $\Phi$ then the diagonal pairing
 $<\psi_i, \phi_{-j-1}>_{diag} = \d_{i,j}$ in terms of
 generating functions looks like
 \[
 <\psi(u),\phi(x)>_{diag}=\frac{1}{u-x},\hsp
\mid x\mid \ll 1\ll \mid u\mid
 \]
 Let $A$ be linear operator in the space $\Phi$. If we use the notation
 $<\psi(u), A\phi(x)>_{diag}$ as a pairing of $\psi(u)$ and
 $A\phi(x)$ under the condition that the pairing of $\psi(u)$ and
 $\phi(x)$ is known to be diagonal then (\ref{4.16}) can be read as
 \bn
 <\der K_+(u), K_-(x)>= <\der K_+(u),
(T-T^{-1})K_-(x)>_{diag} \label{4.17}
 \ed
 where
 $T:Tf(x)=f(x+1)$ is a shift operator. We have from
 (\ref{4.17}):
 \[
 <\der K_+(u),(T-T^{-1})^{-1} K_-(x)>=
 <\der K_+(u), K_-(x)>_{diag}=\frac{1}{u-x}
 \]
 Now we can formally invert an operator $(T-T^{-1})$
   as
 \bn
 (T-T^{-1})^{-1}=-T-T^3-T^5-\ldots
 \label{4.17a}
 \ed
 which gives the diagonalization of bilinear form (\ref{4.15}):
 \bn
 \sum_{n\geq 0}<-\der K_+(u),K_-(x+1+2n)>=\frac{1}{u-x}
 \label{4.18}
 \ed

 {\bf Remark.} The inverse (\ref{4.17a}) to a difference derivative
  $T-T^{-1}$ is only right-inverse operator (just as usual integral).
  We can define it in a different manner like
 \bn
 (T-T^{-1})^{-1}=T^{-1}+T^{-3}+T^{-5}-\ldots
 \label{4.17b}
 \ed
  for instance and obtain a diagonalization form
 \bn
 \sum_{n\geq 0}<\der K_+(u),K_-(x-1-2n)>=\frac{1}{u-x}
 \label{4.18a}
 \ed
  Both formulas work for proper regions of finite-dimensional
  representations of ${\cal D}Y(g)$.

   Using the notations $(\psi(u))_i = \psi_i$ for $\psi(u)=
  \sum\psi_iu^{-i-1}$ and
  $$\Res \psi(u)\otimes\phi(x)=\sum_i\psi_i
 \otimes \phi_{-i-1}$$
  we interprete (\ref{4.18}) as the following expression of the
  factor $R_H$ of the universal $R$-matrix for ${\cal D}Y(sl_2)$:
\[
R_H=\prod_{n\geq 0}\exp\sum(-\der
K_+(u))_i\otimes(K_-(x+2n+1))_{-i-1}
  \]
\bn
=\prod_{n\geq 0}\exp\Res(-\der
K_+(u)i\otimes K_-(x+2n+1))
\label{4.20}
\ed
or if we use (\ref{4.18a}),
\[
R_H=\prod_{n\geq 0}\exp\sum_{i\geq 0}(\der
K_+(u))_i\otimes(K_-(x-2n-1))_{-i-1}
  \]
\bn
=\prod_{n\geq 0}\exp\Res(\der
K_+(u)\otimes K_-(x-2n-1))
\label{4.20a}
\ed
We can summarize the calculations of the universal
$R$-matrix for ${\cal D}Y(sl_2)$ in the following theorem.
\begin{theorem}
Let $K_\pm(u)=\log h_\pm(u)$. Then the universal
$R$-matrix for ${\cal D}Y(sl_2)$ may be presented in
factorized form
\bn
R=R_ER_HR_F
\label{4.21}
\ed
where
\bn
R_E=\; \prod_{i\geq 0}^{\rightarrow}
\exp (-e_i\otimes f_{-i-1}),
\label{4.21a}
\ed
\bn
R_H=\prod_{n\geq 0}\exp\Res(-\der
K_+(u)\otimes K_-(x+2n+1)),
\label{4.21b}
\ed
\bn
R_F=\; \prod_{i\geq 0}^{\leftarrow}
\exp (-f_i\otimes e_{-i-1}).
\label{4.21c}
\ed
\end{theorem}
Let us return to the general case. Just as for
$sl_2$-case, we have the following description of the
pairing (\ref{4.15})
 in terms of the derivative of $K_{i,+}(u)$:
\bn
<\der K_{i,+}(u),K_{j,-}(x)>=
\frac{1}{u-x+\frac{1}{2}(\a_i,\a_j)}-
\frac{1}{u-x-\frac{1}{2}(\a_i,\a_j)}.
\label{4.22}
\ed
It is more convenient to collect fields $K_{i,\pm}(u)$
to  vector valued generating functions
\[
\bar{K}_\pm(u)=\left(
\begin{array}{c}
K_{1,\pm}(u)\\
K_{2,\pm}(u)\\
\ldots\\
K_{r,\pm}(u)
\end{array}
\right) \hsp r={\rm rank}\, g.
\]
In terms of vector valued fields $\bar{\psi}(u)$ and
$\bar{\phi}(x)$ the diagonal pairing $<\psi_{i,n},
\phi_{j,-m-1}>= \d_{i,j} \d_{n,m}>$ looks like
\[
<\bar{\psi}(u),\bar{\phi}(x)>_{diag}=\frac{E}{u-x}
\]
where $E$ is $r\times r$ identity matrix.

Let now $B$ be a symmetrized Cartan matrix of $g$
  with matrix elements being integers without common
divisor, $B_{i,j}= (\a_i, \a_j)$ and $B(q)$ be a
$q$-analog of $B$:  \bn
B_{i,j}(q)=[(\a_i,\a_j)]_q=\frac{q^{(\a_i,\a_j)}-
q^{-(\a_i,\a_j)}}{q-q^{-1}}
\label{A(q)}
\ed
Here we use standard notation $[a]_q=
 \frac{q^a-q^{-a}}{q-q^{-1}}$.

 Let $D(q)$ be an inverse matrix to $B(q)$. One can
see that $D(q)$ can be presented in a form \bn
 D(q)=\frac{1}{[l(g)]_q}C(q)
 \label{C(q)}
 \ed
 where $C(q)$ is a matrix with matrix coefficients $C_{i,j}(q)$ being
 polynomials of $q$ and of $q^{-1}$ with positive
 integer coefficients and $l(g)$ being positive
 integer.  Actually the calculation of $\det\,B(q)$
 show that $l(g)$ is proportional to a dual Coxeter
 number $\h$ of $g$ (see also Table
 (\ref{table})below:  \[ l(g)=\h\;\;\;\mbox{for}\:
g=A_n, E_6, E_7, E_8,\hsp l(g)=2\h\;\;\;\mbox{for}\:
 g=B_n, D_n, F_4, \] \bn l(g)=3\h\;\;\;\mbox{for}\:
 g=G_2,\hsp l(g)=4\h\;\;\;\;\mbox{for}\:g= C_n
 \label{l(g)} \ed In these notations the pairing
 (\ref{4.22}) can be written as \bn <\der
 \bar{K}_+(u),K_-(x)>= <\der \bar{K}_+(u),\left(
(q-q^{-1})B(q)\right)\mid_{q=T^{\frac{1}{2}}}K_-(x)>_{diag}
 \label{4.23}
 \ed
 where a shift operator $T:Tf(x)=f(x+1)$ is substituted inside
 rhs of (\ref{4.23}) instead of $q^2$. Next we deduce that
 \bn
  <\der \bar{K}_+(u),
 (T^{\frac{l(g)}{2}}-T^{-\frac{l(g)}{2}})^{-1}C(T^\frac{1}{2})K_-(x)>=
 \frac{E}{u-x}.
 \label{4.24}
 \ed
 Returning to original notations we get the following diagonalization
 of the pairing (\ref{4.15}):
 \bn
 \sum_{n\geq 0}<-\der
 K_{i,+}(u),\sum_lC_{l,j}(T^{\frac{1}{2}})
 K_{l,-}(x+(n+\frac{1}{2})l(g))>=\frac{\d_{i,j}}{u-x}.
 \label{4.25}
 \ed
 Let us note once more again that,just as in  $sl_2$
case (\ref{4.20a}) it is possible to write down
another diagonalization of the pairing
(\ref{4.15}) for any $g$.  We can summarize the
calculations in the following theorem.
\begin{theorem}
 \label{th4.4}
 Let $K_{i,\pm}(u)= \log
h_{i,\pm}(u)$, $B$ be a symmetrized Cartan matrix of
$g$ with matrix elements being integers without common
divisor, $B_{i,j}= (\a_i, \a_j)$ and $B(q)$ be a
$q$-analog of $B$.  Let $D(q)$ be an inverse matrix to
 $B(q)$, and $C(q)$ be defined by the relation $C(q)=
 [l(g)]_qD(q)$, $l(g)$ from (\ref{l(g)}),
$C_{i,j}(q)\in \mbox{{\bf Z}}[q,q^{-1}]$.

Then the factor $R_H$ of the universal R-matrix for ${\cal D}Y(g)$ can
be presented as
 \[
 R_H=\prod_{n\geq 0}\exp -\sum_{i,j=1,\ldots,r}
 \sum_{m\geq 0} \left(\der K_{i,+}(u)\right)_m
 \otimes \left(C_{j,i}(T^{\frac{1}{2}})
 K_{j,-}(x+(n+\frac{1}{2})l(g))\right)_{-m-1}=
 \]
 \bn
 =\prod_{n\geq 0}\exp -\sum_{i,j=1,\ldots ,r}\Res
 \left(\der K_{i,+}(u) \otimes
 C_{j,i}(T^{\frac{1}{2}})
 K_{j,-}(x+(n+\frac{1}{2})l(g))\right) .
 \label{4.26}
 \ed
 \end{theorem}

 In order to complete the description of the universal $R$-matrix for
 ${\cal D}Y(G)$  we should extend the
 description of their factors $R_E$ and $R_H$ from $sl_2$ case
 to the case of arbitrary simple Lie algebra $g$. Let us first change
 the notations for generators of ${\cal D}Y(g)$. Instead of $e_{i,n}$ we
 use  $e_{\a_i+n\d}$ and instead of $f_{i,n}$ we use
 $e_{-\a_i+n\d}$. Denote also by $\hat{\D}_{Re}$ the set of all real
 roots of corresponding affine nontwissted Lie algebra. Let $\Xi$ be a
 subset of  $\hat{\D}_{Re}$. Recall that a total linear ordering $<$ of
 $\Xi$ is called {\it normal} (or convex) \cite{T} if for any three
 roots $\a ,\b ,\g \in \Xi$, $\g =\a +\b$ we have $\a <\g < \b$ or
 $\b < \g < \b$.

 Let $\Xi_E$ and $\Xi_F$ be the following subsets of $\hat{\D}_{Re}$:
 \bn
 \Xi_E=\{ \g +n\d\;\mid\g\in\D_+(g),n\geq 0\} ,\hsp
 \Xi_F=\{-\g +n\d\;\mid\g\in\D_+(g),n\geq 0\} ,\hsp
 \label{4.27}
 \ed
 Here $\d$ is a minimal imaginary root of $\hat{g}$. Let us equip
 $\Xi_E$ and $\Xi_F$ with two arbitrary  normal
 orderings ${<}_{E}$ and ${<}_{F}$ satisfying
the additional constraint
\bn
 \g +n\d {<}_{E}\;
 \g +m\d ,\;\;\;\mbox{and}\;\;\; -\g +n\d \;
 {>}_{F}\; -\g +m\d \hsp \mbox{if}\;\; n>m
 \label{4.28}
 \ed
 for any $\g\in\D_+(g)$.
 We can define the ''root vectors'' $e_{\pm\nu}$, $\nu\in \Xi_E\cup
  \Xi_F$ by induction following the instruction \cite{KT3}
   \bn
  e_{\nu_3}=[e_{\nu_1},e_{\nu_2}], \hsp
  e_{-\nu_3}=[e_{-\nu_2},e_{-\nu_1}]
  \label{4.29}
  \ed
  if  $\nu_1 <\nu_3 <\nu_2$ and
   $(\nu_1,\nu_2)$ is a minimal segment in a sense of chosen
  orderings containing $\nu_3$, and $e_{\nu_1}$,
 $e_{\nu_2}$ are already being constructed.
 Analogously to the case $U_q(\hat{g})$ \cite{KT3},
 \cite{bonn} one can prove that the procedure
  (\ref{4.29}) is correctly defined and the monomials
  $$ e_{\nu_1}^{n_1}e_{\nu_2}^{n_2} \cdots
  e_{\nu_k}^{n_k},\;\;\;\;
  \nu_1{<}_{E}\,\nu_2{<}_{E}\ldots
 {<}_{E}\nu_k, \;\;\;\; \nu_i\in \Xi_E$$ form a
  basis of subalgebra $E\subset Y(g)$ and the
  monomials
$$ e_{\nu_1}^{n_1}e_{\nu_2}^{n_2} \cdots
  e_{\nu_k}^{n_k},\;\;\;\;
  \nu_1{<}_{F}\nu_2{<}_{F}\ldots
  {<}_{F}\nu_k, \;\;\;\; \nu_i\in \Xi_F$$
form a basis of subalgebra $F\subset Y(g)$.

  Various arguments \cite{KST}, \cite{KT1} show that
  the factors $R_E$ and $R_F$ of the universal
  $R$-matrix for ${\cal D}Y(g)$ should have the
  following form which we state here as a conjecture
  since we did not check the rigorous proof.

  {\bf Conjecture.}  The factors $R_E$ and $R_F$ of the
  universal $R$-matrix for ${\cal D}Y(g)$ have the following form
  \bn
  R_E=\prod_{\vu\in\Xi_E}^{\rightarrow}\exp\left( -a(\vu)e_\vu
  \otimes e_{-\vu}\right) ,
  \label{4.30}
  \ed
  \bn
  R_F=\prod_{\vu\in\Xi_F}^{\rightarrow}\exp\left(
 -a(\vu)e_\vu \otimes e_{-\vu}\right) \label{4.31} \ed
  where the products in (\ref{4.30}) and in (\ref{4.31}) are taken in
  given normal orderings
  ${<}_{E}$ and ${<}_{F}$ satisfying
 (\ref{4.30}).  Normalizing constants $a(\vu)$ are
  taken from the relations
  $$[e_\vu,e_{-\vu}]=\frac{h_\g}{a(\vu)} \;\;\;\;
  \mbox{if}\;\; \vu =\g +n\d\in\Xi_E, \;\;\;\;
  \g\in\D_+(g),$$
  $$[e_{-\vu},e_{\vu}]=\frac{h_\g}{a(\vu)} \;\;\;\;
  \mbox{if}\;\; \vu =-\g +n\d\in\Xi_F, \;\;\;\;
\g\in\D_+(g),$$.
\setcounter{equation}{0}
\section{$R$-matrix for tensor products of evaluation representations
of $Y(sl_2)$}
In this section we demonstrate for the examples of
evaluation representations of $Y(sl_2)$ how the
general formulas (\ref{4.21}), (\ref{4.26}),
 (\ref{4.30}), (\ref{4.31})  forr the universal
 $R$-matrix work in concrete representations of the yangian $Y(sl_2)$.
 Analogous calculations for $U_q(\widehat{sl}_2)$ are presented in
 \cite{KST}.

 One can easily check that an assignment
 \bn
 \phi :
 e_0\rightarrow e,\;\;
 h_0\rightarrow h,\;\;
 f_0\rightarrow f,\;\;\;\;\;\;\;\;
 e_1\rightarrow \frac{h-1}{2}e,\;\;
 f_1\rightarrow f\frac{h-1}{2},
 \label{5.1}
 \ed
 extends to an epimorphism of algebras $Y(sl_2)
\rightarrow U(sl_2)$.  In terms of generating
 functions morphism $\phi$ can be written as \[ \phi
 e_+(u)=\left( u-\frac{h-1}{2}\right)^{-1}e,\hsp \phi
 f_+(u)=f\left( u-\frac{h-1}{2}\right)^{-1},
 \]
 \bn \phi
 h_+(u)=1+\left( u-\frac{h-1}{2}\right)^{-1}ef-
 \left( u-\frac{h+1}{2}\right)^{-1}fe,\hsp \mid u\mid
 \gg 1.  \label{5.2} \ed Morphism $\phi$ is also
 properly defined for ${\cal D}Y(sl_2)$ since one can
 interprete rhs of (\ref{5.2}) as an expansion near zero:
 \[
 \phi e_-(u)=\left( u-\frac{h-1}{2}\right)^{-1}e,\hsp
 \phi f_-(u)=f\left( u-\frac{h-1}{2}\right)^{-1},
 \]
 \bn
 \phi h_-(u)=1+\left( u-\frac{h-1}{2}\right)^{-1}ef-
 \left( u-\frac{h+1}{2}\right)^{-1}fe,\hsp \mid u\mid
 \ll 1.  \label{5.3} \ed Let $V_\l$ and $V_\mu$ be two
 representations of Lie algebra $sl_2$ with highest
 weights $\l$ and $\mu$ (Verma modules or their
 finite-dimensional quotions, for instance), let $V_\l(a)$ and
 $V_\mu(b)$ be corresponding evaluation representations. In concrete
 example when $V_n$ is finite-dimensional representation (dim$V_n=n+1$)
 with highest weight vector $v_0$ and a basis $v_0,v_1,\ldots ,v_n$ we
 have for $V_n(a)$ by (\ref{5.2}):
 \[
 e_iv_k=k\left( a+\frac{n-2k+1}{2}\right)^iv_{k-1}
 \]
 \bn
 f_iv_k=(n-k)\left( a+\frac{n-2k-1}{2}\right)^iv_{k+1}
 \label{5.4}
 \ed
 \[
 h_iv_k= \left(
 (k+1)(n-k)\left( a+\frac{n-2k-1}{2}\right)^i-
 k(n-k+1)
 \left( a+\frac{n-2k+1}{2}\right)^i \right)v_{k}
 \]
 For the calculation of $R$-matrix in $V_\l(a) \otimes V_\mu(b)$ it is
 sufficient to compute $(\phi\otimes \phi) (T_a \otimes T_b)R$ as a
 function of $a-b$ with values in $U(sl_2) \otimes U(sl_2)$. Here $T_d$
 is a shift operator in ${\cal D}Y(sl_2)$, $T_de_\pm(u)= e_\pm(u-d)$,
 $T_dh_\pm(u)= h_\pm(u-d)$, $T_df_\pm(u)= f_\pm(u-d)$. We do this first
 for the factors $R_E$ and $R_F$. Let
 \bn
 y=a-b+\frac{h\otimes 1-1\otimes h}{2}..
 \label{y}
 \ed
 We think of $y$ as of diagonal matrix acting in $V_\l(a)\otimes
 V_\mu(b)$. Substitution of (\ref{5.2}) into (\ref{4.4}) gives the
 following answer:
 \[
 (\phi\otimes\phi)(T_a\otimes T_b)R_E=
\]
\bn
=1+\frac{1}{y-1}e\otimes f+
 \frac{1}{2(y-1)(y-2)}e^2\otimes f^2+\ldots +\frac{1}{n!(y-1)\cdots
 (y-n)}e^n\otimes f^n+\ldots
 \label{5.5}
 \ed
 One can consider rhs of (\ref{5.5}) as a difference analog of ordered
 exponential:
\[
(\phi\otimes\phi)(T_a\otimes T_b)(R_E)=:\exp
(y-1)^{-1}\cdot e\otimes f:_{T^{-1}}
\]
where $T$ is again a shift operator and
\[
:\exp f(y)\cdot g(y):_{T^{\pm 1}}=
1+f(y)\cdot g(y)+\frac{1}{2}
f(y)f(y\pm 1)\cdot g(y)g(y\pm 1)+\ldots
\]
\[
\ldots +
\frac{1}{n!}
f(y)f(y\pm 1)\ldots f(y\pm (n-1))\cdot
g(y)g(y\pm 1)\ldots g(y\pm (n-1))+\ldots
\]

Analogously,
\[
(\phi\otimes\phi)(T_a\otimes T_b)(R_F)=:\exp
(y+1)^{-1}\cdot f\otimes e:_{T}=
\]
\bn
=1+\frac{1}{y+1}e\otimes f+
 \frac{1}{2(y+1)(y+2)}f^2\otimes e^2+\ldots
+\frac{1}{n!(y+1)\cdots (y+n)}f^n\otimes e^n+\ldots
 \label{5.6}
 \ed

 Note that for any weight vector of $V_\l\otimes
 V_\mu$ the series (\ref{5.5}) and (\ref{5.6}) is
 finite and have a form of operator $U(sl_2)\otimes
 U(sl_2)$ with rational coefficients.

 The calculation of
$(\phi\otimes\phi)(T_a\otimes T_b)(R_H)$
 is more complicated. We perform this calculation
 directly in $V_\l\otimes V_\mu$ for simplicity. We
 can rewrite the action of $h_+(u)$ in
 $V_\l(a)$ (\ref{5.4}) as
 \bn
 h_+(u)=
 \frac{(u-a-\frac{\l +1}{2}) (u-a+\frac{\l +1}{2})}
 {(u-a-\frac{h +1}{2}) (u-a-\frac{h -1}{2})},
 \hsp \mid u\mid \gg 1
 \label{5.7}
 \ed
 and analogously the action of $h_-(x)$ in
 $V_\mu(b)$  as
 \bn
 h_-(x)=
 \frac{(x-b-\frac{\mu +1}{2}) (x-b+\frac{\mu +1}{2})}
 {(x-b-\frac{h +1}{2}) (x-b-\frac{h -1}{2})},
 \hsp \mid x\mid \ll 1.
 \label{5.8}
 \ed

 We get from (\ref{5.7}) and (\ref{5.8}):
 \[
 \der K_+(u)=d\log h_+(u)=
\]
\bn
 =\frac {1}{(u-a-\frac{\l +1}{2})}+
 \frac {1}{(u-a+\frac{\l +1}{2})}-
 \frac {1}{(u-a-\frac{h +1}{2})}-
 \frac {1}{(u-a-\frac{h -1}{2})},
 \;\;\; \mid u\mid \gg 1,
 \label{5.9}
 \ed
 and
 \bn
 K_-(x)=
 \log \frac{(x-b-\frac{\mu +1}{2}) (x-b+\frac{\mu
 +1}{2})} {(x-b-\frac{h +1}{2}) (x-b-\frac{h -1}{2})},
 \hsp \mid x\mid \ll 1.
 \label{5.9a}
 \ed  \medskip
 The rest of computations can be easily performed
 using the followiing Lemma, where we assume, as
 usually, that $\mid x\mid \ll 1\ll\mid u\mid$.
 \begin{lemma}
 \bn
 \prod_{n\geq 0}\exp\Res
 \left(
 \frac{1}{u-\g}\cdot \log\frac{x-\a +2n+1}{x-\b
 +2n+1}\right) =
 \frac{\Gamma \left(\frac{\g -\b +1}{2}\right)}
 {\Gamma \left(\frac{\g -\a +1}{2}\right)} .
 \label{5.11}
 \ed
 \end{lemma}
 Denoting $h_1=h\otimes 1$, $h_2=1\otimes h$,
 $ c=\frac{a-b}{2}$ we get finally the following
 horrible answer:

 \[
 R_H\mid_{V_\l(a)\otimes V_\mu(b)}=
 \frac{\Gamma\left( c+\frac{\l
 -\mu}{4}+\frac{1}{2}\right)
 \Gamma\left( c-\frac{\l
 -\mu}{4}+\frac{1}{2})\right)
\Gamma\left( c+\frac{\l
 +\mu}{4}+1\right)
 \Gamma\left( c-\frac{\l +\mu}{4}\right)}
 {\Gamma\left( c+\frac{\l -h_2}{4}+\frac{1}{2}
 \right)
 \Gamma\left( c-\frac{\l +h_2}{4}+\frac{1}{2}
 \right)
 \Gamma\left( c+\frac{\l +h_2}{4}+1\right)
 \Gamma\left( c-\frac{\l +h_2}{4}\right)}
 \cdot
 \]

 \bn
  \cdot
 \frac{\Gamma\left( c+\frac{h_1-h_2}{4}\right)
 \left(\Gamma\left( c+\frac{h_1-h_2}{4}+\frac{1}{2}
 \right)
 \right)^2
 \Gamma\left( c+\frac{h_1 -h_2}{4}+1\right)}
 {\Gamma\left( c+\frac{h_1 -\mu}{4}+\frac{1}{2}
 \right)
 \Gamma\left( c+\frac{h_1 +\mu}{4}+\frac{1}{2}
 \right)
 \Gamma\left( c+\frac{h_1+\mu}{4}+1\right)
 \Gamma\left( c+\frac{h_1-\mu}{4}\right)}
 \label{5.12}
 \ed \medskip

 If we suppose  $\l$ and $\mu$ to be integers and
 $V_\l$, $V_\mu$ be finite dimensional (
 $\mbox{dim}\ V_\l =\l +1$,
 $\mbox{dim}\ V_\mu =\mu +1$,
 then the whole $R$-matrix has rational coefficients
 up to a scalar factor. We can easily find this factor
 $\rho(\l ,\mu)$ normalizing $R$-matrix in such a way
 that a matrix coefficient of tensor product of
 highest weight vector to itself is equal to one. This
 gives
 \bn
 \rho(\l ,\mu)=\frac{
 \Gamma\left( \frac{a-b}{2}+\frac{\l
 +\mu}{4}+\frac{1}{2}\right)
 \Gamma\left( \frac{a-b}{2}-\frac{\l
 +\mu}{4}+\frac{1}{2}\right)}
 {\Gamma\left( \frac{a-b}{2}+\frac{\l
 -\mu}{4}+\frac{1}{2}\right)
 \Gamma\left( \frac{a-b}{2}+\frac{\mu
 -\l}{4}+\frac{1}{2}\right)}
 \label{5.13}
 \ed \medskip

In the next section we describe this scalar factor for
arbitrary finite-dimensional representations of
$Y(g)$.

\setcounter{equation}{0}
\section{The characters of the universal $R$-matrix
and bilinear form on the weights of $h_+(u)$}
Let $V$ and $W$ be two irreducible finite-dimensional
representations of the yangian $Y(g)$. It can be
proved by fusion procedure and by studying of the
$R$-matrix for fundamental representations of the
yangian that the $R$-matrix $R_{V,W}$ intertwining two
coproducts $\D$ and ${\D}'$ in $V(a)\otimes V(b)$ can
be presented as a matrix with rational coefficients of
$a-b$ (here $V(a)$ and $W(b)$ are obtained from $V$
and $W$ by means of a natural shift automorphism of
$Y(g)$). On the other hand, one can apply
$\rho_{V(a)}\otimes \rho_{W(b)}$ to the universal
$R$-matrix. The results should differ by a scalar
phase factor ${<V,W>}$ which appears due to
nonlinear conditions on the universal $R$-matrix:
\bn
(\D\otimes id)R=R^{13}R^{23},
\label{6.1}
\ed
\bn
(id\otimes\D)R=R^{13}R^{12}.
\label{6.2}
\ed
The factor ${<V,W>}$ is by definition unique
modulo rational functions of $(a-b)$ and plays an
important role in scattering theory.

Unfortunately, the intriguing theory of
finite-dimensional representations of yangians is too
young and does not says much about representations:
there is a classification of irreducible modules but
their structure is almost unknown (including the
dimensions and characters). Nevertheless the theory of
highest weight developed by Drinfeld \cite{Dnew} (see
also \cite{Ta1}, \cite{Ta2}) coupled with our
description of the universal $R$-matrix allows to
compute the factor ${<V,W>}$ for arbitrary
irreducible finite-dimensional representations of
yangian $Y(g)$.

Let us recall the basic definitions of highest weight
polynomials of finite-dimensional representation of
$Y(g)$.
\begin{definition}
Let $V$ be a $Y(g)$-module. A vector
$v\in V$ is a highest weight vector if

(i)$\hsp e_{i,+}(u)v=0, \hsp i=1,\ldots ,r$

(ii)$\hsp h_{i,+}(u)v=H_{i}(u)v\;\;\;$ (i.e.,$v$ is an
eigenvector for all $h_{i,n}, n\geq 0$).
\end{definition}
The functions $H_i(u)$, $i=1,\ldots r$ are called
eigenfunctions of $v$.
\begin{theorem} \cite{D83}.

a) Any irreducible finite-dimensional $Y(g)$-module is
generated by a highest weight vector;

b) An irreducible finite-dimensional $Y(g)$-module $V$
with a highest weight vector $v$ is finite-dimensional
iff the eigenfunctions $H_i(u)$ of $v$ can be
presented as ratio of polynomials
\bn
H_i(u)=\frac{P_i(u+\frac{1}{2}(\a_i ,\a_i))}{P_i(u)}
\label{6.3}\ed
\end{theorem}
\begin{definition}
a) The polynomials $P_i(u)$,  $i=1, \ldots ,r$
 defined by the condition (\ref{6.3}) are called
highest weight polynomials of a highest weight vector
$v$ (and of finite-dimensional representation $V$);

b) Finite-dimensional representation of $Y(g)$ with
highest weight polynomials $P_i(u)= u-a, P_j(u)=1,
j\neq i$ is called $i$-th fundamental representation
of $Y(g)$.

We denote $i$-th fundamental representation of $Y(g)$
by $\o_i(a)$.
\end{definition}
Bycomponent multiplication of weight polynomials
endows the set $E(g)$ of all irreducible
finite-dimensional representations of $Y(g)$ with a
structure of abelian (multiplicative) semigroup
generated by fundamental representations. An element
$V$ of $E(g)$ can be presented as
\bn
V=\o_1(a_{1,1})\cdots\o_1(a_{1,i_1})\cdot\ldots\cdot
\o_r(a_{r,1})\cdots\o_r(a_{r,i_r})
\label{6.4}
\ed
which means that highest weight polynomials of $V$
are  $P_j(u)=(u-a_{j,1})\cdots (u-a_{j,i_j})$ and $V$
can be realized as subfactor of tensor product of
fundamental representations $\o (a_{i,j})$ containing
tensor product of highest weight vectors of
 fundamental representations.  Analogously, the
 multiplication law $U=V\cdot W$ implies that an
 irreducible module $U$ is a subfactor of $V\otimes W$ generated by
 image of tensor product of highest weight vectors of $V$ and $W$.

Let now $V$ and $W$ be finite-dimensional irreducible representation of
$Y(g)$ generated by highest weight vectors $v$ and $w$. They can be
endowed wih a structure of $Y_-(g)$ (and of ${\cal D}Y(g)$)-module just by
 reexpansion of matrix coefficients of $e_{i,+}(u)$, $h_{i,+}(u)$,
 $f_{i,+}(u)$, in $u=0$. The general structure (\ref{4.1}) of the universal
 $R$-matrix shows that $v\otimes u$ is an eigenvector of $R$. This gives a
 possibility to normalize scalar phase ${<V,W>}$ for all highest weight
 representations of $Y(g)$.
 \begin{definition}
 A scalar function ${<V,W>}$ defined by the condition
 \bn
 R(v\otimes w)={<V,W>} v\otimes w
 \label{6.6}
 \ed
 where $R$ is universal $R$-matrix for ${\cal D}Y(g)$, $v$ and $w$
 are highest  weight vectors of $V$ and $W$, is called the character of
 the universal R-matrix $R$ corresponding to highest weight representations
 $V$ and $W$.
 \end{definition}
 The following lemma explaines that the character of the universal
 $R$-matrix may be considired as (multiplcative) bilinear form on $E(g)$.
 \begin{lemma}
  \label{lemma6.1}
 Let $V,V_1,V_2;\;\; W,W_1,W_2\;\in E(g)$. Then
 \bn
 {<V_1\cdot V_2,W>}={<V_1,W>}\cdot {<V_2,W>},
 \label{6.7}
 \ed
 \bn
 {<V,W_1\cdot W_2>}={<V,W_1>}\cdot {<V,W_2>}
 \label{6.7a}
 \ed
 \end{lemma}
 The proof immediately follows from (\ref{6.1}), (\ref{6.2}).

 The general expresion (\ref{4.26}) for the factor $R_H$ of the universal
 $R$-matrix  allows to find out the character of $R$ for arbitrary
 finite-dimensional representations of $Y(g)$. Indeed, the action of $R$ on
 tensor product $v\otimes w$ of highest weight vectors
 reduces to the action of its factor $R_H$. If
 $P_i(u)$ are highest weight polynomials of $v$ and
  $Q_j(u)$ are highest weight polynomials of $w$ then the action of the
  fields $\der K_{i,+}(u)$ on $v$ is given by the expression
  \[
  \der K_i(u)v= \left( d\log P_i
  \left( u+\frac{1}{2}(\a_i,\a_i)\right) -
  d\log P_i(u)\right) v, \hsp\mid u\mid \gg 1;
  \]
  $K_{-,j}(x)$ act on $w$ as
  \[
  K_{j,-}(x)w=\log \frac{Q_j(x+\frac{1}{2}(\a_i,\a_j))}
  {Q_j(x)}w,\hsp\mid x\mid \ll 1.
  \]
  The rest is technical application of (\ref{4.26}). Due to Lemma
  \ref{lemma6.1}
  it is sufficient to compute the characters, corresponding to
  tensor product of fundamental representations.

  Let us recall the notations of previous section.
 Now  again $B$ is a symmetrized Cartan matrix of $g$
 with matrix elements being integers without common
 divisor, $B_{i,j}= (\a_i, \a_j)$ $i,j-1,\ldots ,r$
 and $B(q)$ is a $q$-analog of $B$; $D(q)$is an
 inverse matrix to $B(q)$ and $C(q)$ is a $r\times r$
   matrix with coeffitients from $\mbox{{\bf Z}}[q,q^{-1}]$ defined by the
   condition
 $D(q)=\frac{1}{[l(g)]_q}C(q)$, where $l(g)$ is proportional to dual
 Coxeter  number $\h (g)$ (see (\ref{l(g)})). A presentation (\ref{C(q)}),
 (\ref{l(g)}) of the inverse to $q$-analogue of
symmetrized Cartan matrix follows from calculation of
 $\mbox{det}B(q)$:
 \bn
 \begin{array}{cccc}
 g\;\;\;\;\;\;&\det  B(q)\;\;&\h(g)&l(g)\\\
  \; &\; &\; &\; \\
 A_l& [l+1]_q&l+1&l+1\\
B_l&\frac{\textstyle{[2(2l-1)]_q}}
{\textstyle{[2]_q[2l-1]_q}}&2l-1&2(2l-1)\\
C_l&\frac{\textstyle{[2]_q[2(l+1)]_q}}
{\textstyle{[l+1]_q}}&l+1&4(l+1)\\
 D_l&\frac{\textstyle{[2]_q[2l-2)]_q}}
{\textstyle{[l-1]_q}}&2l-2&2(2l-2)\\
E_6&\frac{\textstyle{[2]_q[3]_q[12]_q}}
{\textstyle{[4]_q[6]_q}}&12&12\\
E_7&\frac{\textstyle{[2]_q[3]_q[18]_q}}
{\textstyle{[6]_q[9]_q}}&18&18\\
E_8&\frac{\textstyle{[2]_q[3]_q[5]_q[30]_q}}
{\textstyle{[6]_q[10]_q[15]_q}}
&30&30\\
 F_4&\frac{\textstyle{[2]_q[3]_q[18]_q}}
{\textstyle{[6]_q[9]_q}}&9&18\\
G_2&\frac{\textstyle{[2]_q[3]_q[12]_q}}
{\textstyle{[4]_q[6]_q}}&4&12
\end{array}
\label{table}
\ed
 The calculations with $R_H$ gives the following
 \begin{theorem}
 Let $C_{i,j}(q)=\sum_k C_{i,j}^kq^k$, $C_{i,j}^k\in \mbox{{\bf Z}}_+$.
 The character $<\o_i(a),\o_j(b)>$ of universal $R$-matrix,
 corresponding to fundamental representations $\o_i(a)$ and $\o_j(b)$ is
 equal to
 \[
 <\o_i(a),\o_j(b)>=
 \]
 \bn
 \prod_k\left(\frac{
 \Gamma\left(\frac{a-b}{l(g)}+\frac{l(g)-k}{2l(g)}\right)
 \Gamma\left(\frac{a-b}{l(g)}+\frac{l(g)-k+(\a_j,\a_j)-(\a_i,\a_i)}
 {2l(g)}\right)}
 {\Gamma\left(\frac{a-b}{l(g)}+\frac{l(g)-k-(\a_i,\a_i)}
 {2l(g)}\right)
 \Gamma\left(\frac{a-b}{l(g)}+\frac{l(g)-k+(\a_j,\a_j)}
 {2l(g)}\right)}
 \right)^{C_{i,j}^k}
 \label{6.8}
 \ed
 \label{th6.1}
 \end{theorem}
 For instance, for $Y(sl_2)$ the pairing (\ref{6.8})
   looks like \bn <\o(a),\o(b)>=
 \frac{\Gamma\left(\frac{a-b}{2}+\frac{1}{2}\right)^2}
 {\Gamma\left(\frac{a-b}{2}\right)
 \Gamma\left(\frac{a-b}{2}+1\right)}
 \label{6.9}
 \ed
 and, more generally,
 \bn
 <\prod_i\o(a_i),\prod_j\o(b_j)>=\prod_{i,j}
 \frac{\Gamma\left(\frac{a_i-b_j}{2}+\frac{1}{2}\right)^2}
 {\Gamma\left(\frac{a_i-b_j}{2}\right)
\Gamma\left(\frac{a_i-b_j}{2}+1\right)}
 \label{6.10}
 \ed
 which agrees with (\ref{5.13}) since $V_n(a)= \o(a- \frac{n-1}{2})
  \o(a- \frac{n-3}{2}) \cdots \o(a+ \frac{n-1}{2})$
in $E(sl_2)$.

Quasiclassically
 $R(u)= 1+ \frac{\Omega}{u}$ and quasiclassical limit of the form
  (\ref{6.6}) on highest weights of evaluation
  representation  $V_\l(a)$ and $V_\mu(b)$ of Lie
  algebra $g[t]$ should be $\frac{<\l ,\mu >}{a-b}$
  where $<,>$ is invariant scalar product in $h^*$
  ($h$ is Cartan subalgebra of $g$).

%% We can also
%formally symmetrize form (\ref{6.5}) %%putting
%$\nu^S_{<V,W>}= \nu_{<V,W>} \nu_{<W,V>}$.  %%  Then
%$\nu^S_{<W,V>} $ is a deformation of a %%   natural
%scalar product on highest weights of %%   $V_\l(a)$
%and $V_\mu(b)$ equal to $<\l ,\mu>\d %%   (a-b)$.

   It will be interesting to
   obtain combinatorial and geometric interpretation
of the form (\ref{6.6}).

\end{document}